\documentclass[aps,pra,notitlepage,twocolumn]{revtex4-1}
\usepackage{amsmath,amssymb}
\usepackage{dsfont}
\usepackage{graphicx}

\begin{document}

\newcommand{\bra}[1]    {\langle #1|}
\newcommand{\ket}[1]    {| #1\rangle}
\newcommand{\tr}[1]    {{\rm Tr}\left[ #1 \right]}
\newcommand{\av}[1]    {\left\langle #1 \right\rangle}
\newcommand{\proj}[1]{\ket{#1}\bra{#1}}

\title{Role of Particle Entanglement in the Violation of Bell Inequalities}

\author{T. Wasak$^1$, A. Smerzi$^2$ and J. Chwede\'nczuk$^1$}
\affiliation{$^1$Faculty of Physics, University of Warsaw, ul. Pasteura 5, PL--02--093 Warszawa, Poland\\
$^2$QSTAR, INO-CNR and LENS,  Largo Enrico Fermi 2, 50125 Firenze, Italy}

\begin{abstract}
  Entanglement between two separate systems is a necessary resource to violate a Bell inequality in a test of local realism. 
  We demonstrate that to overcome the Bell bound, this correlation must be accompanied by the entanglement
  between the constituent particles. This happens whenever a super-selection rule imposes a constraint on feasible local operations.
  As we show in an example, the necessary particle entanglement might solely result from their indistinguishability.
  Our result reveals a fundamental relation between the non-locality and the particle entanglement.
\end{abstract}

\maketitle

The ``spooky action at the distance'' stands out among the most striking consequences of quantum mechanics \cite{epr}. 
This term was coined by Albert Einstein to
underline how counterintuitive it is that a seemingly local manipulation on one part of a system immediately 
affects its other distant part without any transfer of physical information.
Such an effect contradicts the postulates of the ``local realism'':
Two quantum spin-$\frac12$ particles,
if prepared in an entangled state and sent into distant regions $A$ and $B$, cannot be treated as individual objects up until the measurements are made.
Operations performed in $A$ and $B$, though local in space, act globally on the system.
The non-locality of quantum mechanics can be quantified by a series of inequalities---first considered by Bell in \cite{bell}---
for the correlations between the outcomes of local measurements
\cite{bell_rmp,chsh,test1,test2,test3,test4,test4,test5,test6,test7,test8,test9,test10,test11,test12,ent_rmp,banaszek1}.
While the violation of the Bell inequalities was first observed decades ago \cite{test2,test3,test4,test4},
only recently a loophole-free deviation from local realism has been demonstrated experimentally \cite{loophole, loophole2}.

Not all $A$-$B$ entangled states violate a known Bell inequality \cite{werner}, 
but it has been known since long that all {\it pure} states do violate a Bell inequality  \cite{gisin}. 
For illustration, consider a pure state $\ket\psi$  shared between $A$ and $B$, which decomposed into the localized states reads
\begin{equation}\label{sch}
  \ket{\psi}=\sum_ic_i\ket{\phi_i}_A\otimes\ket{\chi_i}_B.
\end{equation}
If the state is $A$-$B$ entangled---which happens when at least two coefficients of this expansion, say $c_i$ and $c_{i'}$, are non-zero---
a Bell inequality \cite{gisin} will be violated, by 
locally coupling $\ket{\phi_i}_A$ with $\ket{\phi_{i'}}_A$ and $\ket{\chi_i}_B$ with $\ket{\chi_{i'}}_B$. These couplings
probe the quantum coherence between $\ket{\phi_i}_A\otimes\ket{\chi_i}_B$ and $\ket{\phi_{i'}}_A\otimes\ket{\chi_{i'}}_B$.

However, sometimes local operations and/or measurements are prohibited by some superselection rule (SSR). The SSR
is a restriction imposed on quantum mechanics forbidding coherences between eigenstates of certain observables \cite{ssr_wick,ssr2}. 
For the purpose of this manuscript, the SSR can be formulated as follows:
local operations/measurements cannot create/detect coherences between states with different number of particles of a given type. 
Here, a particle is understood as a discrete object, indistinguishable from other of the same kind, and carrying a set of fundamental quantum numbers, such as the 
charge or the baryon and lepton numbers \cite{ssr_ent1}.

To illustrate the impact of the SSR on the feasible local operations, consider
two states localized in $A$: one which contains a single sodium 23 atom, denoted by $\ket{{\rm ^{23}Na}}_A$ and the other with a rubidium 87 atom, denoted by $\ket{{\rm ^{87}Rb}}_A$.
Although these states have the same number of atoms, the atoms are different.
Any operation or measurement coupling these states would not preserve the number of atoms of a given kind or---from another perspective---would not conserve the number of protons, electrons and neutrons.
Therefore, such coupling is forbidden by the SSR \cite{ssr_wick,ssr2}. 
Another known example is a single particle coherently distributed
among $A$ and $B$ \cite{enk}, namely
\begin{equation}\label{pure}
  \ket\psi=\frac1{\sqrt2}\Big(\ket1_A\otimes\ket0_B+\ket0_A\otimes\ket1_B\Big).
\end{equation}
The SSR formulated above prohibits the creation or the detection of a superposition
of the vacuum $\ket0_A$ with the state containing one particle $\ket1_A$,
thus the local operations
cannot create/detect quantum coherences between the
two components of $\ket\psi$.
From the point of view of physically realizable local operations one can effectively replace the pure state (\ref{pure}) with an incoherent mixture
\begin{eqnarray}
  \ket\psi\bra\psi\rightarrow\hat\varrho_{\rm eff}=\frac1{2}&&\Big(\ket1\bra1_A\otimes\ket0\bra0_B\nonumber\\
  &&+\ket0\bra0_A\otimes\ket1\bra1_B\Big).\label{replace}
\end{eqnarray}
Although the state (\ref{pure}) is $A$-$B$ entangled,
due to the SSR the resulting $\hat\varrho_{\rm eff}$ is $A$-$B$ separable (i.e., non-entangled) and as such does not violate any Bell inequality \cite{ssr_ent1,ssr_ent2,ssr_ent3}. 
Note that for photons, to which the SSR does not apply, the local coupling of 
$\ket0_A$ with $\ket1_A$ is allowed and indeed the state (\ref{pure}) violates a Bell inequality \cite{loophole, loophole2,enk}.

Inspired by this example we formulate and prove a general theorem:
the restriction imposed on the local operations by the SSR cast all particle-separable states
to be effectively $A$-$B$ separable. In other words, in presence of SSR not only the $A$-$B$ entanglement
but also
the entanglement of
particles shared by $A$ and $B$ are necessary for the violation of any Bell inequality.
We demonstrate that this latter resource might origin solely from the particle indistinguishability \cite{identical_plenio}.

To set the stage and proceed with the proof, we note that the $A$-$B$
separable states have a general form
\begin{equation}
  \hat\varrho=\sum_i\,p_i\,\hat\varrho^{(i)}_A\otimes\hat\varrho^{(i)}_B,\label{sepAB}
\end{equation}
where $p_i$'s are the statistical weights.
This relation is established by the decomposition of the total Hilbert space into the local sub-spaces
\begin{equation}\label{prodAB}
  \mathcal H=\mathcal H_A\otimes\mathcal H_B.
\end{equation}
We will demonstrate that in presence of SSR and in the context of Bell inequalities,
the quantum state should also be inspected through the decomposition of $\mathcal H$ into the product of single-particle subspaces
\begin{equation}\label{prodN}
  \mathcal H=\bigotimes_{i=1}^N\mathcal H_i.
\end{equation}
Here $N$ is a number of particles shared by $A$ and $B$.
In analogy to Eq.~(\ref{sepAB}), particle-separable states are
\begin{equation}\label{sep}
    \hat\varrho=\sum_i\,p_i\,\hat\varrho^{(1)}_i\otimes\ldots\otimes\hat\varrho^{(N)}_i,
\end{equation}
and particle-entangled are those that cannot be expressed in this way.
We adapt this general formula to the $A$-$B$ geometry and
show that all such states do not violate any Bell inequality in presence of SSR. This means that the quantum state shared by $A$ and $B$ must necessarily be
particle-entangled to violate any Bell inequality.



%

First, we consider a collection of distinguishable particles. The basic building block of the $N$-body density matrix (\ref{sep}) is the one-body pure state, which for the $i$-th particle reads
\begin{equation}\label{dist1}
  \ket{\psi_i}=\left(\alpha(\psi_i)\,\hat\psi_{_i}^{(A)\dagger}+\beta(\psi_i)\,\hat\psi_{i}^{(B)\dagger}\right)\ket{0}.
\end{equation}
Here $\hat\psi_{i}^{(k)\dagger}$ creates a quantum of a field associated with this particle in the region $k$, $\ket{0}$ is the vacuum and $|\alpha(\psi_i)|^2+|\beta(\psi_i)|^2=1$.
According to Eq.~(\ref{sep}), the density matrix of $N$ particles forming a separable state is an incoherent mixture of the one-body matrices.
Since one must allow each particle to be distributed among the regions in every way---and this for the $i$-th body is governed by the field $\psi_i$--- the Eq.~(\ref{sep})
translates into
\begin{equation}\label{dens_dist}
  \hat\varrho=\int\!\!{\mathcal D}\psi_1\cdots\int\!\!{\mathcal D}\psi_N\,\mathcal{P}(\psi_1,\ldots,\psi_N)\bigotimes_{i=1}^N\ket{\psi_{i}}\bra{\psi_{i}}.
\end{equation}
Here, the joint probability $\mathcal{P}(\psi_1,\ldots,\psi_N)$ determines the partition of all the bodies among $A$ and $B$. The symbol ${\mathcal D}\psi_i$ is the integration measure
over the set of fields $\psi_i$.

Note that the product $\otimes$ in Eq.~(\ref{dens_dist}) refers to the decomposition of the Hilbert space as in Eq.~(\ref{prodN}).
On the other hand, if we decomposed $\hat\varrho$ into the states residing in $\mathcal H_A$ and $\mathcal H_B$, the state would not be
$A$-$B$ separable---it could not be written in the form of Eq.~(\ref{sepAB}).
This is because the density matrix (\ref{dens_dist}) has multiple terms which account for the $A$-$B$ entanglement, due to the inter-region coherence of each component from Eq.~(\ref{dist1}).
However, the particles form a separable state, therefore to identify the feasible local operations in presence of SSR, each single-particle state can be considered separately. 
Following the example from Eq.~(\ref{pure}), the SSR enforces every $\ket{\psi_{i}}\bra{\psi_{i}}$ to be replaced with 
\begin{eqnarray}\label{single}
  \ket{\psi_{i}}\bra{\psi_{i}}\rightarrow\hat\varrho_{\rm eff}(\psi_{i})&=&|\alpha(\psi_i)|^2\hat\psi_{i}^{(A)\dagger}\ket{0}\bra{0}\,\hat\psi_{i}^{(A)}\nonumber\\
  &+&|\beta(\psi_i)|^2\hat\psi_{i}^{(B)\dagger}\ket{0}\bra{0}\,\hat\psi_{i}^{(B)}.
\end{eqnarray}
This expression, plugged back into (\ref{dens_dist}) gives
\begin{equation}\label{eff_d}
  \hat\varrho_{\rm eff}=\int\!\!{\mathcal D}\psi_1\cdots\int\!\!{\mathcal D}\psi_N\,\mathcal{P}(\psi_1,\ldots,\psi_N)\bigotimes_{i=1}^N\hat\varrho_{\rm eff}(\psi_i).
\end{equation}
Since the inter-region coherence is washed out already on the single-particle level of Eq.~(\ref{single}) and the integral over the fields does not introduce any quantum coherence,
the effective $N$-body density matrix $\hat\varrho_{\rm eff}$ is both particle- and $A$-$B$-separable (for the rigorous proof, see the
Appendix). To conclude, the SSR transform the state (\ref{dens_dist}) into (\ref{eff_d}), which has the form of Eq.~(\ref{sepAB}), and as such it will not violate any Bell inequality.

If the distinguishable particles are entangled, violation of some Bell inequality in presence of SSR might be possible. For illustration, 
consider an electron ($e$) and a proton ($p$) forming a particle- and  $A$-$B$-entangled state
\begin{equation}
  \ket\psi=\frac1{\sqrt{2}}\Big(\ket{\uparrow_e}_A\otimes\ket{\uparrow_p}_B+\ket{\downarrow_e}_A\otimes\ket{\downarrow_p}_B\Big),
\end{equation}
where the arrows denote the projection of the spin of each particle. Now, local operations can be executed by coupling
$\ket{\uparrow_e}_A$ with $\ket{\downarrow_e}_A$ and $\ket{\uparrow_p}_B$ with $\ket{\downarrow_p}_B$. Therefore, according to the discussion below Eq.~(\ref{sch}), this state will violate a Bell inequality. 
On the other hand, take an alternative particle- and  $A$-$B$-entangled state
\begin{equation}
  \ket\psi=\frac1{\sqrt{2}}\Big(\ket{\uparrow_e,\uparrow_p}_A\otimes\ket{0}_B+\ket{0}_A\otimes\ket{\downarrow_e,\downarrow_p}_B\Big).
\end{equation}
It will not violate any Bell inequality, because SSR forbid the coupling of $\ket{\uparrow_e,\uparrow_p}_A$ with $\ket{0}_A$ and
$\ket{\downarrow_e,\downarrow_p}_B$ with $\ket{0}_B$. This second example highlights the fact that when the SSR apply, both the particle and the $A$-$B$ entanglement are only necessary, but not
sufficient to drive the violation of a Bell inequality.

We now turn to bosons, for which the equivalent of Eq.~(\ref{dens_dist}) is \cite{cauchy,cauchy_long}
\begin{equation}\label{sep_bos_rho}
  \hat\varrho=\int\!\!{\mathcal D}\psi\,\mathcal{P}(\psi)\ \ket{\psi}\bra{\psi}.
\end{equation}
Here $\ket{\psi}$ is the spin coherent state, which reads
\begin{equation}\label{sep_bos}
  \ket{\psi}=\left(\alpha(\psi)\,\hat\psi^{(A)^\dagger}+\beta(\psi)\,\hat\psi^{(B)^\dagger}\right)^N\ket0.
\end{equation}
The language of the second quantization allows to immediately identify the relation between the regions $A$ and $B$, i.e, provides the state decomposed according to Eq.~(\ref{prodAB}).
This can be seen by writing Eq.~(\ref{sep_bos}) in terms of $A/B$  occupation states, i.e,
\begin{equation}\label{sep_bos2}
  \ket{\psi}=\sum_{n_\psi=0}^NC_{n_\psi}\ket{n_\psi}_A\otimes\ket{N-n_\psi}_B,
\end{equation}
where $C_{n}=\sqrt{\binom N{n}}\alpha(\psi)^n\beta(\psi)^{N-n}$.
This expression plugged into Eq.~(\ref{sep_bos_rho}) gives
\begin{eqnarray}
    \hat\varrho&=&\int\!\!{\mathcal D}\psi\,\mathcal{P}(\psi)\sum_{n_\psi=0}^N\sum_{m_\psi=0}^NC^*_{n_\psi}C_{m_\psi}\times\nonumber\\
    &\times&\ket{n_\psi}\bra{m_\psi}_A\otimes\ket{N-n_\psi}\bra{N-m_\psi}_B\label{dens_bos}.
\end{eqnarray}
In presence of SSR, local operations cannot couple $\ket n_k$ with $\ket{n'}_k$. In this context, the state (\ref{sep_bos2}) can be effectively replaced by
\begin{eqnarray}
  \ket{\psi}\bra\psi\rightarrow\hat\varrho_{\rm eff}(\psi)=\sum_{n_\psi=0}^N&&|C_{n_\psi}|^2\ket{n_\psi}\bra{n_\psi}_A\otimes\nonumber\\
  &&\otimes\ket{N-n_\psi}\bra{N-n_\psi}_B,
\end{eqnarray}
which is both particle- and $A$-$B$-separable. Also the effective density matrix
\begin{equation}
  \hat\varrho_{\rm eff}=\int\!{\mathcal D}\psi\,\mathcal{P}(\psi)\hat\varrho_{\rm eff}(\psi)
\end{equation}
is $A$-$B$ separable. Thus for bosons, in presence of SSR the particle entanglement
is a necessary resource for the violation of any Bell inequality.

We now show that the entanglement extracted solely from the indistinguishability of bosons might be sufficient for the
violation of the Bell inequality.
\begin{figure}[t!]
  \includegraphics[clip, scale=0.6]{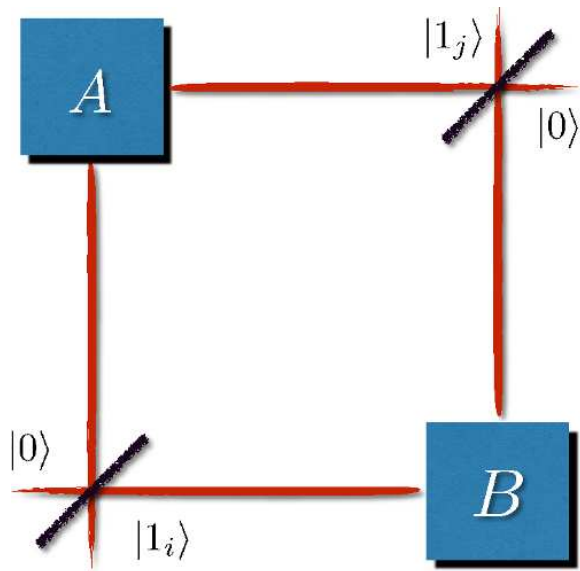}
  \caption{(color online) Particles of type $i$ (bottom left) and of type $j$ (top right) are coherently split and sent into the regions $A$ and $B$.
    If they are distinguishable, the system will not violate any Bell inequality in presence of SSR, because they form a particle-separable state. 
    To contrary, identical particles in this configuration form a particle-entangled state due to their indistinguishability. In such case, the violation of some Bell inequality is possible.}
  \label{scheme_yurke}
\end{figure}
Consider a particle of type $i$ and a particle of type $j$ in a state
\begin{equation}\label{ex1}
  \ket\psi=\ket{1_i}\otimes\ket{1_j}
\end{equation}
entering the system through the two ports \cite{yurke}, shown in
Fig.~\ref{scheme_yurke}. The two beam-splitters distribute the signal among $A$ and $B$, giving
\begin{equation}\label{ex2}
  \ket\psi=\frac12\Big(\ket{1_i}_A+\ket{1_i}_B\Big)\otimes\Big(\ket{1_j}_A+\ket{1_j}_B\Big).
\end{equation}
The symbol $\otimes$ in Equations (\ref{ex1}) and (\ref{ex2}) multiplies the single-particle states, therefore it refers to the decomposition of the Hilbert space as in Eq.~(\ref{prodN}). 
To analyze the relation between $A$ and $B$, we switch to the second-quantization by expanding the product
and expressing $\ket\psi$ in terms of $A$- and $B$-occupation states. For instance, $\ket{1_i}_A\otimes\ket{1_j}_B\rightarrow\ket{1_i,0_j}_A\otimes\ket{0_i,1_j}_B$, giving
\begin{eqnarray}
  \ket\psi&=&\frac12\Big(\ket{1_i,0_j}_A\otimes\ket{0_i,1_j}_B+\ket{0_i,1_j}_A\otimes\ket{1_i,0_j}_B\nonumber\\
  &+&\ket{1_i,1_j}_A\otimes\ket{0,0}_B+\ket{0,0}_A\otimes\ket{1_i,1_j}_B\Big).    \label{dist_bs_b}
\end{eqnarray}
Now, the product relates to the decomposition of the Hilbert space into $\mathcal H_A$ and $\mathcal H_B$, as in Eq.~(\ref{prodAB}) and clearly the state is $A$-$B$ entangled.
If the particles are distinguishable, i.e., $i\neq j$, 
they are not entangled and the only pair of states in $A$ with equal number of
particles are $\ket{1_i,0_j}_A$ and $\ket{0_i,1_j}_A$ (and analogically in $B$). These states cannot be locally
coupled in presence of SSR and the system will not violate any Bell inequality.
On the other hand, if the particles are identical, i.e., $i=j$,
the state (\ref{ex1}) is particle-entangled state due to the indistinguishability. Now,
the coupling of $\ket{1_i,0}_k$ with $\ket{0,1_{j=i}}_k$
can be realized and the $A$-$B$ entanglement, together with particle entanglement coming solely from indistinguishability \cite{identical_plenio},
will drive the violation of some Bell inequality.

If a separable state contains a group of bosons and a group of distinguishable particles, all above arguments can be applied to each sup-group separately,
because local operations, in presence of SSR prohibit the transmutation of a particle of one type into another.
Moreover, if the state reveals incoherent particle-number fluctuations, that are consistent with SSR, each fixed-$N$ sector can be considered separately,
leading to the same conclusion---particle-separable states do not reveal non-classicality in any Bell test.
Also, one could extend the system by adding an auxiliary reference frame to the particle-separable state \cite{banaszek2,dowling}. 
A composite system is created and it undergoes local operations in $A$ and $B$.
If this reference frame is a quantum system, to which the SSR applies, then according to our proof, as long as this extension does not introduce any particle entanglement,
the composite system will remain effectively $A$-$B$ separable.

%

Finally, we point that our result is in line with what is known in quantum interferometry \cite{giovannetti2004quantum,varenna}. 
There, a collection of particles passes through the two arms of an interferometer.
During the propagation, a phase is imprinted on one of the arms. In order to surpass the shot-noise limit for the
sensitivity of the phase estimation at the output, during the phase-imprint two conditions
must be satisfied: the two arms must be entangled (which corresponds to the $A$-$B$ entanglement in our case), and the system must be particle-entangled.

To summarize, we have shown that in presence of super-selection rules, mode entanglement must be accompanied by
entanglement between the particles in order to violate a Bell inequality.
Our proof applies to any system, where the particle-entangled/non-entangled dichotomy is present. 
This is the case of distinguishable particles, bosons, or systems where bosons and distinguishable
particles co-exist. Naturally, fermions form only the particle entangled states due to the Pauli principle.
Our result puts the particle entanglement on par with the $A$-$B$ mode entanglement, 
as a necessary condition for the violation of the local realism. We have demonstrated that the particle entanglement necessary for the violation of the Bell inequalities might result solely from 
the indistinguishability of bosons.

T. W. acknowledges the support of the Ministry of Science and Higher Education programme ``Iuventus Plus'' for years 2015-2017, project number IP2014 050073.

\section*{Appendix}

The one-body pure state for the particle of type $i$, which is distributed among the regions $A$ and $B$ reads
\begin{equation}\label{dist1.s}
  \ket{\psi_{i}}=\left(\alpha(\psi_i)\hat\psi_{i}^{(A)^\dagger}+\beta(\psi_i)\hat\psi_{i}^{(B)^\dagger}\right)\ket{0}.
\end{equation}
We introduce a shortened notation, where
\begin{equation}
  \alpha(\psi_i)\hat\psi_{i}^{(A)^\dagger}\equiv\hat\Phi_{i}^{(A)^\dagger}\ \ \ \mathrm{and}\ \ \ \beta(\psi_i)\hat\psi_{i}^{(B)^\dagger}\equiv\hat\Phi_{i}^{(B)^\dagger}
\end{equation}
With this at hand, the state (\ref{dist1.s}) is
\begin{equation}\label{dist2.s}
  \ket{\psi_{i}}=\sum_{\kappa_i\in\{A,B\}}\hat\Phi_{i}^{(\kappa_i)^\dagger}\ket{0}
\end{equation}
Every density matrix of $N$ particles in a separable state can be expressed as
\begin{equation}\label{dens_dist.s}
  \hat\varrho=\int\!\!{\mathcal D}\psi_1\cdots\int\!\!{\mathcal D}\psi_N\,\mathcal{P}(\psi_1,\ldots,\psi_N)\bigotimes_{i=1}^N\ket{\psi_{i}}\bra{\psi_{i}},
\end{equation}
where $\mathcal{P}(\psi_1,\ldots,\psi_N)$ is a probability distribution.
We now insert the expression (\ref{dist2.s}) into (\ref{dens_dist.s}) and obtain
\begin{widetext}
  \begin{equation}
    \hat\varrho=
    \sum_{\kappa_1\in\{A,B\}}\cdots\sum_{\kappa_N\in\{A,B\}}\
    \sum_{\kappa'_1\in\{A,B\}}\cdots\sum_{\kappa'_N\in\{A,B\}}
    \int\!\!{\mathcal D}\psi_1\cdots\int\!\!{\mathcal D}\psi_N\,\mathcal{P}(\psi_1,\ldots,\psi_N)\hat\Phi_{1}^{(\kappa_1)^\dagger}\cdots\hat\Phi_{N}^{(\kappa_N)^\dagger}\ket{0}\bra{0}
    \hat\Phi_{1}^{(\kappa'_1)}\cdots\hat\Phi_{N}^{(\kappa'_N)}.
  \end{equation}
\end{widetext}
In this state, the quantum correlation between the regions $A$ and $B$ arises from the one-body coherence, which is
represented in the independent sums over $\kappa_i$ and $\kappa'_i$.
The restriction imposed on local operations require that in each regions, only states with a fixed number of particles of each type can couple. This means that the sums over
$\kappa_i$ and $\kappa'_i$ effectively do not run independently, and the state reduces to
\begin{widetext}
  \begin{equation}
    \hat\varrho_{\rm eff}=
    \sum_{\kappa_1\in\{A,B\}}\cdots\sum_{\kappa_N\in\{A,B\}}\
    \int\!\!{\mathcal D}\psi_1\cdots\int\!\!{\mathcal D}\psi_N\,\mathcal{P}(\psi_1,\ldots,\psi_N)\hat\Phi_{1}^{(\kappa_1)^\dagger}\cdots\hat\Phi_{N}^{(\kappa_N)^\dagger}\ket{0}\bra{0}
    \hat\Phi_{1}^{(\kappa_1)}\cdots\hat\Phi_{N}^{(\kappa_N)}.
  \end{equation}
\end{widetext}
This state does not reveal any quantum coherence and is $A$-$B$ separable. $\Box$

\end{document}